\documentclass[twocolumn,british,report,floatfix]{revtex4}
\usepackage{amssymb,amsmath} 
\usepackage{color}
\usepackage{graphicx}

\begin{document}

\title{A Protocol For Cooling and Controlling Composite Systems by Local Interactions} 

\author{Daniel Burgarth$^{1}$ and Vittorio Giovannetti$^{2}$}

\affiliation{$^{1}$Computer Science Department, ETH Z\"urich, CH-8092 Z\"urich,
Switzerland\\
$^{2}$NEST-CNR-INFM \& Scuola Normale Superiore, piazza dei Cavalieri
7, I-56126 Pisa, Italy}

\begin{abstract}
We discuss an explicit protocol which allows one to externally cool and control
a composite system by operating on a small subset of it. 
 The scheme permits to transfer \emph{arbitrary and unknown} quantum states from a memory
on the network ({}``upload  access'') as well as the inverse
({}``download access''). In particular it yields a method for cooling the
system. 
\end{abstract}

\maketitle

\section{Introduction}
Repetitive applications of the same quantum transformation 
have been exploited to achieve noise protection~\cite{VIOLA3},
cooling, state preparation~\cite{HOMOGENIZATION1,WELLENS,YUASA1}, 
and  quantum state transfer~\cite{MEMORYSWAP}. 
Motivated by the above results, in Ref.~\cite{PAPER} we
developed a scheme for
controlling larger systems when control is only assumed to be available
on a subsystem. Once this is achieved, 
apart from cooling and state preparation, 
it is also possible to perform
arbitrary quantum data processing (e.g. measurements, unitary rotations).
This is similar in spirit to
universal quantum interfaces of Ref.~\cite{SETH}, 
but our approach allows us to specify explicit protocols
and to give lower bounds for fidelities. 
These techniques are also related with the ``asymptotic completeness'' property introduced by
K\"ummerer and Maassen~\cite{Kummerer,WELLENS} which allows one to
control a system by coupling it with quantum mediators.

In the present paper we review the scheme of Ref.~\cite{PAPER}
by showing how arbitrary quantum
states can be written into (i.e. prepared on) a large system, and read
from it, by \emph{local} control only. This implies that arbitrary quantum
operations on the system state can be performed.
An important specific task is the cooling of the system to its ground state. Using some heuristic argument, we will provide an estimate of the convergence time of the cooling and we will test it with
 some numerical examples. 
We develop the protocol in several steps. First, 
we show that the system of interest can be actively
brought to its ground state by replacing its controlled 
part with fresh ``cold'' qubits from a memory.
We then find that cooling implies that the 
information about the initial system state is transferred
into the memory, and design a linear map that decodes this information. 
Since this map is generally not unitary, we use the polar decomposition to find 
its best unitary approximation.  The fidelity of information 
decoding can then be lower bounded by the overlap of the system state with
its ground state. Finally we design the reverse operation 
allowing us to transfer information from the memory to the system.

The material is organized as follows. In Sec.~\ref{sec:Protocol}
the protocol is presented in its general lines.
In Sec.~\ref{sec:Coding-transformation} we give a detailed
derivation of the coding and decoding transformations and derive
bounds for the fidelities.
Numerical estimations of the protocol performances are given in Sec.~\ref{sec:cool}
focusing on the case of locally controlled Heisenberg-like coupled spin networks.
Conclusion and remarks are in Sec.~\ref{sec:con} while technical material
is presented in the Appendices.

\section{The Protocol\label{sec:Protocol}}

\begin{figure}[t]
\begin{centering}\includegraphics[width=.8\columnwidth]{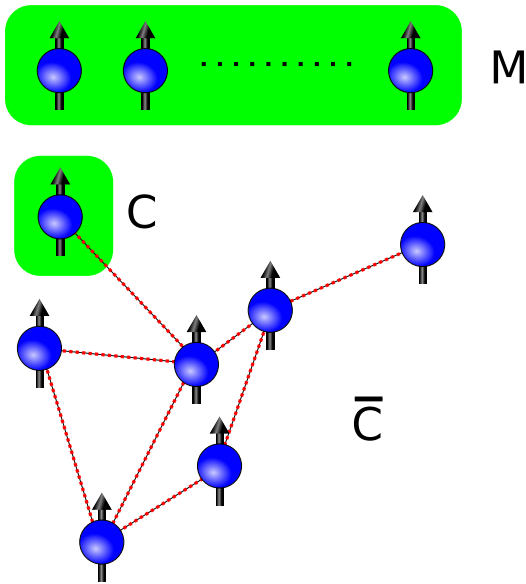}\par\end{centering}
\caption{\label{fig:memory} Example of the model discussed in the text. Here the system $C\bar{C}$ is formed by 7 spins characterized by some
time-independent Hamiltonian $H$ (the coupling are represented by red lines connecting the spins). The system $C\bar{C}$ can only be controlled
by acting on a (small) subsystem $C$ (in this case represented by the
uppermost spin of the network). 
The coupling $H$ can - in some cases - \emph{mediate} the local control
on $C$ to the full system $C\bar{C}.$ In our case, system $C$ is
controlled by performing regular swap operations $S_{\ell}$ between
it and a 2-dim quantum memory $M_{\ell}.$ }
\end{figure}

Consider a  composed system  
 described by
the Hilbert space 
$\mathcal{H}=\mathcal{H}_{C}\otimes\mathcal{H}_{\bar{C}}\otimes\mathcal{H}_{M}.$ 
We assume that full control (the ability to prepare states and apply
unitary transformations) is possible on system $C$ and $M,$ but
no control is available on system $\bar{C}.$ Moreover, we assume that
$C$ and $\bar{C}$ are coupled by a time-independent Hamiltonian
$H.$ We show here that under certain assumptions, if the system $C\bar{C}$
is initialized in some arbitrary state we can transfer (``download'')
this state into the system $M$ by applying some operations between
$M$ and $C$ only. Likewise, by initializing the system $M$ in the correct
state, we can transfer (``upload'') arbitrary states on the system
$C\bar{C.}$ The system $M$ functions as a \emph{quantum memory}
and must be at least as large as the system $C\bar{C}$. As sketched
in Fig.~\ref{fig:memory} we can imagine the memory to be split into sectors
$M_{\ell},$ having the same dimension
of $C$, i.e. 
$\mathcal{H}_{M}=\bigotimes_{\ell=1}^{L}\mathcal{H}_{M_{\ell}}$
with 
$\textrm{dim}\mathcal{H}_{M_{\ell}}=\textrm{dim}\mathcal{H}_{C}$.

%%%%%%%%%%%%%%%%%%%%%%%%%%%%%%%%%%%%%%%%%%%%%%%%%%%%%%%%%%%%%%%%%%%%%%%
%%%%%%%%%%%%%%%%%%%%%%%%%%%%%%%%%%%%%%%%%%%%%%%%%%%%%%%%%%%%%%%%%%%%%%
\subsection{Downloading info from $C\bar{C}$ to $M$}\label{s:down}

The  downloading protocol we present here is composed by two stages:
a {\em swapping} stage, in which at regular time intervals  
we couple the subsystem $C$ to the first $L$ 
memories $M$; and a {\em decoding}  stage
in which we apply a unitary transformation to the first $L$ memories in order
to recover the initial state of $C\bar{C}$.
As we will see for any finite $L$ our analysis does not guarantee that 
 the fidelity  between the recovered state and the initial
state of $C\bar{C}$ is perfect. However, in Sec.~\ref{sec:fid} it will be shown that 
by augmenting  $L$ one can make the fidelity arbitrarily close to one.

We assume that the memory is initialized in a factorized 
state of the form \begin{equation}
|0\rangle_{M}\equiv\bigotimes_{\ell=1}^L|0\rangle_{M_{\ell}}\;, 
\end{equation}
where $|0\rangle$ is a state whose properties will be
specified in the following. 
To download  a generic state, we let the system $C\bar{C}$ to 
evolve for a while according to its Hamiltonian $H$, perform
a unitary gate which couples $C$ to one of 
the sectors of  $M$, let $C\bar{C}$ evolve again and so forth. 
More specifically,
at step $\ell$ of the protocol we perform a unitary swap $S_\ell \equiv 
S_{C M_\ell}$
between system $C$ and system $M_{\ell}$~\cite{NOTE2}. After the $L$th swap
operation the protocol stops. This is the {\em swapping} stage
and it is characterized by the 
unitary operator 
\begin{equation}
W\equiv S_{L}US_{L-1}U\cdots S_{\ell}U\cdots S_{1}U\;,
\label{www} 
\end{equation}
where  $U=\exp\left[ -iHt\right] $ is the time-evolution 
of $C\bar{C}$ for some fixed time interval $t.$ 
As discussed in Ref.~\cite{PAPER}, the reduced evolution of
the system $\bar{C}$ under the transformation~(\ref{www})  can be expressed in terms
of the following completely positive CP map~\cite{NIELSEN}
 \begin{equation}
\tau(\rho_{\bar{C}})\equiv\textrm{Tr}_{C}\left[U\left(\rho_{\bar{C}}\otimes|0\rangle_{C}\langle0|\right)U^{\dag}\right]\;, \label{CPMAP}
\end{equation}
where $|0\rangle_{C}$ is the state that is swapped in from the memory and 
$\textrm{Tr}_{C} [\cdots] $ indicate the partial trace over the subsystem $C$. 
Indeed, 
after $L$ swaps 
the state of $\bar{C}$ is obtained by taking the partial trace with
respect to $C$ and $M$ of the vector
$W \left( |\psi\rangle_{C\bar{C}} |0\rangle_M  \right)$ where
$|\psi\rangle_{C\bar{C}}$ is the initial state of $C\bar{C}$, i.e.
\begin{eqnarray}
\rho_{\bar{C}}^{(L)} &=&  \mbox{Tr}_{C M} 
\left[ W \left( |\Psi\rangle_{C\bar{C}}\langle \Psi| \otimes
 |0\rangle_M\langle 0|   \right) W^\dag \right] \nonumber \\
\nonumber \\ \label{impo1} 
&=& \underbrace{\tau \circ \tau \circ \cdots \circ 
\tau}_{\mbox{$L-1$ times}} \;
(\rho_{\bar{C}}^{\prime})
\equiv  \tau^{L-1} (\rho_{\bar{C}}^\prime) \;, \label{NEWREF}
\end{eqnarray}
 where ``$\circ$'' represents
the composition  of super-operators~\cite{NIELSEN}
 and 
  \begin{eqnarray} \label{newref1}
 \rho_{\bar{C}}^\prime \equiv  \mbox{Tr}_C \big[ 
U|\psi\rangle_{C\bar{C}}\langle \psi| U^\dag \big] \;,
\end{eqnarray}
(for an explicit
derivation of this expression see Appendix~\ref{appendix1}).  

Our main assumption  is that the map $\tau$ is ergodic with 
pure fixed point which we  denote as $|0\rangle_{\bar{C}}$. 
Explicitly this means that 
 the only state which is left invariant by
$\tau$  is the vector $|0\rangle_{\bar{C}}$, i.e.
\begin{eqnarray}
\tau(\rho_{\bar{C}}) = \rho_{\bar{C}}        \quad 
\Longleftrightarrow \quad 
\rho_{\bar{C}} = |0\rangle_{\bar{C}}\langle 0| \;. \label{condition20}
\end{eqnarray}
As shown in Refs.~\cite{Gohm2004,Giovannetti} this implies
that the channel $\tau$ is relaxing (mixing), that is 
\begin{eqnarray}
\lim_{n \rightarrow \infty} \tau^n(\sigma_{\bar{C}})=|0\rangle_{\bar{C}}\langle 0| \;,
\label{fafa}
\end{eqnarray}
for \emph{all} $\sigma_{\bar{C}}$. This condition gives rise
to the controllability of the system.
Indeed from Eq.~(\ref{fafa}) it follows that 
for sufficiently large $L$, an initial state 
of the form $|\psi\rangle_{C\bar{C}}\otimes|0\rangle_M$ 
can be approximated as 
\begin{eqnarray}
\label{approx}
W \big(
|\psi\rangle_{C\bar{C}}\otimes|0\rangle_M\big)
 \approx |0\rangle_{C\bar{C}} \otimes |\Phi(\psi) \rangle_M.
\end{eqnarray}
The right hand side of this equation factorizes into pure states because
the transformation $W$ is unitary, and because
both the initial state of $C\bar{C}$ and $M$ and the final state
of $C\bar{C}$ are pure.  This implies that, in the asymptotic limit of infinitely many
protocol steps (i.e. $L\gg1$), the system $C\bar{C}$ has been ``cooled'' into the
state $|0\rangle_{C\bar{C}}$ while all the information 
regarding the initial state $|\psi\rangle_{C\bar{C}}$ must be contained in
 the vector $|\Phi(\psi)\rangle_M$~\cite{NOTE3}.
Furthermore, it is at least intuitively clear that such information
can be recovered by the application  of a proper unitary ``decoding'' 
operation $V^\dag$  on $M$ which does not depend on the input state of
the system ({\em decoding} stage), i.e.~\cite{notezero}
\begin{eqnarray} 
V^\dag\; |\Phi(\psi) \rangle_M 
\approx  |\psi\rangle_M \label{hhh}\;.
\end{eqnarray}
At a mathematical level,  the convergence of the downloading protocol described
above only depends upon the invariant property~(\ref{condition20}) --- see Ref.~\cite{PAPER}.
In  Sec.~\ref{sec:Coding-transformation} we will briefly review such a proof and provide a characterization of 
the unitary transformation $V$.
 
\subsection{Uploading info from $M$ to $C\bar{C}$} \label{s:up}

For uploading states on the system $C\bar{C}$, we again make use of the unitarity
of $W.$ Let us again first give a simple hand-waving  argument why this is possible. 

Suppose you want to drive the system into the state $|\psi\rangle_{C\bar{C}}$.
To do this, you first use the downloading protocol 
to make sure that the system is in the state $|0\rangle_{C\bar{C}}$ (``cooling''). 
Then you bring the memories into the state 
$|\Phi(\psi)\rangle_M$ they  would have been ended up
in case one was trying to download $|\psi\rangle_{C\bar{C}}$ from $C\bar{C}$ into $M$ 
as in Eq.~(\ref{approx}).
Now the quantum recurrence theorem~\cite{recurrence} implies that there is a 
$m$ such that 
\begin{eqnarray}
\label{recu}
|\psi\rangle_{C\bar{C}} \otimes |0\rangle_M &\approx& 
W^m \big( |\psi\rangle_{C\bar{C}} \otimes |0\rangle_M\big) 
\nonumber \\
&\approx& W^{m-1} \big( |0\rangle_{C\bar{C}}\otimes |\Phi(\psi)\rangle_M\big)
 \;,
\end{eqnarray}
where we have made use of Eq.~(\ref{approx}). Hence by applying $W$ $m$ times you have approximately initialized $|\psi\rangle_{C\bar{C}}$. 
Of course it remains to be shown that \emph{unknown} states
can be written to the system, too. This and the mathematical details will be discussed in the next section.
Another problem with Eq.~(\ref{recu}) is that the recurrence parameter $m$ typically needs to be \emph{huge}, scaling double exponentially with the number of qubits in the system. There 
are however  alternative, more efficient ways of implementing an uploading process
from $M$ to $C\bar{C}$. The simplest one is of course
to apply the inverse transformation $W^{-1}=W^\dag$ to the state of Eq.~(\ref{approx}). 
Indeed the protocol we presented in Ref.~\cite{PAPER} is based on this idea, 
which is a generalization of~\cite{WELLENS}.
 Unfortunately the inverse of $W$ is generally
unphysical in the sense that it requires backward time evolutions $U^{-1}$,
i.e. one would have to wait \emph{negative} time steps between the swaps
(see however Ref.~\cite{echo} for cases in which such an inverse
time evolution can be  implemented by clever external control techniques).

To overcome this problem we introduce an extra hypothesis. Specifically
we consider the case in which the invariant property~(\ref{condition20})
holds also for the channel $\tau^\prime$ obtained from
Eq.~(\ref{CPMAP}) by replacing $U$ with $U^\dag$, i.e. 
 \begin{equation}
\tau^\prime(\rho_{\bar{C}})\equiv\textrm{Tr}_{C}\left[U^\dag 
\left(\rho_{\bar{C}}\otimes|0\rangle_{C}\langle0|\right)U\right]\;. 
\label{CPMAPprime}
\end{equation}
Under this condition, similarly to the case of $W$ discussed in the previous section,
one can verify that   in the limit of large $L$, {\em i)} the transformation 
\begin{equation}
W^\prime \equiv S_{L}U^\dag S_{L-1}U^\dag \cdots S_{\ell}U^\dag \cdots S_{1}U^\dag\;,
\label{wwwprime} 
\end{equation}
applied to $|\psi\rangle_{C\bar{C}}\otimes|0\rangle_M$ will converge to a vector of the form 
$|0\rangle_{C\bar{C}} \otimes |\Phi^\prime(\psi) \rangle_M$; {\em ii)} 
there exists a unitary transformation $V^\prime$ which does not depend
upon $|\psi\rangle$ and  which applied to $M$ gives
\begin{eqnarray} 
V^{\prime\dag} 
|\Phi^\prime(\psi) \rangle_M  \approx  |\psi\rangle_M \label{hhhprime} \;.
\end{eqnarray}
From this  we can write
\begin{eqnarray} 
&&|\psi\rangle_{C\bar{C}}\otimes|0\rangle_M
\approx (W^\prime)^\dag \; V^\prime \;\big(
 |0\rangle_{C\bar{C}}  \otimes |\psi\rangle_M \big)
\nonumber  \\ \nonumber 
&&\approx (U S_1  \cdots U S_\ell \cdots U S_{L-1} U S_L)\; V^\prime 
\;\big( |0\rangle_{C\bar{C}}  \otimes |\psi\rangle_M \big).
\\  \label{hhhprime1}
\end{eqnarray}
What it is relevant for us is  the fact that now 
the unitary transformation on the input state 
$|0\rangle_{C\bar{C}}  \otimes |\psi\rangle_M$
does not involve ``time-reversal'' evolutions $U^{-1}$ but only
``proper'' time evolution $U$.
Therefore, by  imposing the condition~(\ref{condition20}) on $\tau^\prime$,
we are able to define an uploading protocol which transfers an unknown state $|\psi\rangle$
from $M$ to $C\bar{C}$. Similarly to the downloading scheme it
is composed by two stages:
an {\em encoding} stage in which we apply the unitary transformation $V^\prime$
to  ``prepare'' the memory $M$ and a {\em swapping} stage in which
we apply the unitary
\begin{eqnarray} \label{tildewww}
(W^\prime)^\dag = U S_1 \cdots U S_\ell \cdots U S_{L-1} U S_L\;,
\end{eqnarray}
by recursively coupling $C$ to the $M$ through swaps.

\begin{figure}[t]
\begin{centering}\includegraphics[width=1\columnwidth]{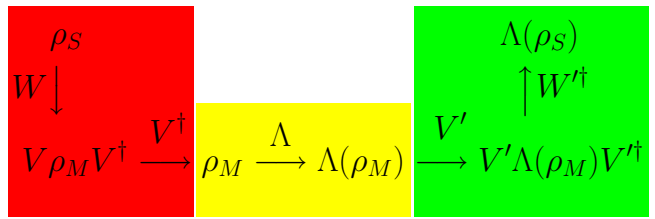}\par\end{centering}
\caption{\label{fig:scheme} Summary of the scheme: any CP map $\Lambda$ can be applied to the system by acting on the memory instead through the transformations shown in the figure. The red and green areas
represent the downloading and uploading part  of the protocol,
respectively.
The unitary operators $W$ and $W^{\prime\dagger}$ of Eqs.~(\ref{www}) and (\ref{tildewww})
are generated by acting on the memory and a small subsystem of the system only; $V^\dag$ and $V^\prime$ are instead the decoding and encoding unitary transformations
introduced in Eqs.~(\ref{hhh}) and (\ref{hhhprime}), 
respectively --- see also Sec.~\ref{sec:Coding-transformation}.}
\end{figure}

Two remarks are mandatory. On one hand, as in the case of the downloading protocol, 
the convergence of the transformation~(\ref{hhhprime1}) only depends upon the 
invariant condition~(\ref{condition20}) of the channel $\tau^\prime$.
On the other hand, there exists a large class
of physically relevant Hamiltonians $H$ 
(e.g. nearest neighbors Heisenberg coupling Hamiltonians)
for which  both $\tau$ and $\tau^\prime$
verify the such  condition --- we refer the reader to Ref.~\cite{PAPER}
for details. For such Hamiltonians, our analysis will yield
both a simple downloading and uploading mechanism.
Putting these two elements together one can also realize more
sophisticated controls. For instance, as shown in Fig.~\ref{fig:scheme},
 one can perform any quantum 
transformation $\Lambda$ on $C\bar{C}$ by first downloading
its state on $M$, transforming it, and finally uploading the 
final state back into the system.

%%%%%%%%%%%%%%%%%%%%%%%%%%%%%%%%%%%%%%%%%%%%%%%%%%%
%%%%%%%%%%%%%%%%%%%%%%%%%%%%%%%%%%%%%%%%%%%%%%%%%

\section{Coding transformation}\label{sec:Coding-transformation}

In this section we derive the decoding 
transformation $V^\dag$ that relates states
on the memories $M$ to the states on $C\bar{C}$ in the downloading
protocol. 
To do so we exploit the formal decomposition of the evolved state of the system
after $L$ steps (see Appendix~\ref{appendix2}).
The encoding transformation $V^\prime$ of the uploading protocol
can be obtained in a similar way.
  
Consider  an orthonormal basis $\left\{ |\psi_{k}\rangle_{C\bar{C}}\right\} $
of $\mathcal{H}_{C\bar{C}}$. According to Eq.~(\ref{eq:main}) after $L$ swaps it becomes 
\begin{eqnarray}
\lefteqn{W \big(|\psi_{k}\rangle_{C\bar{C}}|0\rangle_{M}\big) }\label{eq:maink}\\
 &  & = |0\rangle_{C}\otimes\left[\sqrt{\eta_{k}}|0\rangle_{\bar{C}}|\phi_{k}\rangle_{M}+\sqrt{1-\eta_{k}}|\Delta_{k}\rangle_{\bar{C}M}\right]\;, \nonumber 
\end{eqnarray}
where  $|\Delta_{k}\rangle_{\bar{C}M}$ is a vector orthogonal to $|0\rangle_{\bar{C}}$
 and $\eta_k\approx 1$ as in Eq.~(\ref{eq:fid11}). 
This equation shows that with high probability, the transformation $W$ maps
the orthonormal vectors $|\psi_k\rangle_{C\bar{C}} $ 
into the vectors $|\phi_k\rangle_M$ of the first $L$ memories. For any finite choice of $L$, the latter are 
typically not mutually orthogonal. However one
 can use Eq.~(\ref{eq:fid11}) to show that in the limit of
large $L$ the vectors $|\phi_k\rangle_M$ become approximately 
orthogonal. Indeed 
from the unitarity of $W$ and from Eq.~(\ref{eq:maink}) and (\ref{eq:main2}) we
can establish the following identity \begin{eqnarray}
\lefteqn{_{M}\langle\phi_{k}|\phi_{k'}\rangle_{M}}\label{QQQ}\\
 &  &  = \sqrt{\eta_{k}\;\eta_{k'}}\;\delta_{kk'}+\sqrt{\eta_{k}\;
(1-\eta_{k'})}\;{}_{\bar{C}CM}\langle\psi_{k}0|\tilde{\Delta}_{k'}
\rangle_{\bar{C}CM}\nonumber \\
 &  & +\sqrt{\eta_{k'}\;(1-\eta_{k})}\;{}_{\bar{C}CM}\langle
\tilde{\Delta}_{k}|\psi_{k'}0\rangle_{\bar{C}CM}\nonumber \\
 &  & +\sqrt{(1-\tilde{\eta}_{k})(1-\tilde{\eta}_{k'})}\;{}_{C\bar{C}M}
\langle\tilde{\Delta}_{k}|\tilde{\Delta}_{k'}\rangle_{C\bar{C}M}\;.\nonumber \end{eqnarray}
To simplify this expression we define $\eta_{0}\equiv\min_{k}\eta_{k}$. 
Since 
Eq.~(\ref{eq:fid11}) applies to all $\eta_k$ the parameter $\eta_0$ must satisfy
the inequality
\begin{eqnarray} \label{neweq}
1 -\eta_0  \leqslant  K \;(L-1)^{d_{\bar{C}}}\;\kappa^{L-1}\;.
\end{eqnarray}
Furthermore from Eq.~(\ref{QQQ}) 
it follows that for $k\neq k'$ one has
 \begin{eqnarray}
\lefteqn{|{_{M}\langle}\phi_{k}|\phi_{k'}\rangle_{M}|}\nonumber \\
 &  & \leqslant |\sqrt{\eta_{k}\;(1-\eta_{k'})}\;|{}_{\bar{C}CM}\langle\psi_{k}0|\tilde{\Delta}_{k'}
\rangle_{\bar{C}CM}|\nonumber \\
 &  & +\sqrt{\eta_{k'}\;(1-\eta_{k})}\;|{}_{\bar{C}CM}\langle\tilde{\Delta}_{k}|\psi_{k'}0\rangle_{\bar{C}CM}|\nonumber \\
 &  & +\sqrt{(1-\tilde{\eta}_{k})(1-\tilde{\eta}_{k'})}\;|{}_{C\bar{C}M}\langle\tilde{\Delta}_{k}|\tilde{\Delta}_{k'}\rangle_{C\bar{C}M}|\nonumber \\
 &  & \leqslant 
2\sqrt{1-\eta_{0}}+(1-\eta_{0})\;\leqslant 
\;3\sqrt{1-\eta_{0}}\label{IMPRT} \;,
\end{eqnarray}
which according to Eq.~(\ref{neweq}) and using the fact 
that the parameter $\kappa$ is strictly smaller than $1$,
shows that for large $L$ the
vectors $|\phi_{k}\rangle_{M}$ and
$|\phi_{k'}\rangle_{M}$  become orthogonal.

Define then the linear operator $D$ on ${\cal H}_{M}$ which performs the
following transformation \begin{eqnarray}
D|\psi_{k}\rangle_{M}=|\phi_{k}\rangle_{M}\;,\label{DEFD}\end{eqnarray}
with $|\psi_{k}\rangle_{M}$ being orthonormal vectors of $M$ which
represent the states $\left\{ |\psi_{k}\rangle_{C\bar{C}}\right\} $
of $\mathcal{H}_{C\bar{C}}$. Formally they are obtained by a partial
isometry from $\bar{C}C$ to $M$ and are ``good'' representations of the $|\psi_k\rangle_{C\bar{C}}$. 
The operator $D$ in some sense ``corrects'' the non-orthogonality of the $|\phi_k\rangle_M$:
indeed its inverse (when  definable) 
allows us to pass from these
approximate images of the $|\psi_k\rangle_{C\bar{C}}$ 
to the good representations $|\psi_k\rangle_M$. Therefore $D^{-1}$ seems to be a good candidate
for defining our decoding transformation $V$.
Unfortunately however $D$ is NOT unitary 
(it maps an orthonormal set of states into a non-orthonormal one) and typically will not
be even invertible.

The idea is then to replace $D$ with its 
\emph{best unitary approximation} $V$~\cite[p 432]{HORNJOHNSON}. The latter 
is obtained by taking the polar  decomposition of $D$, i.e. 
\begin{eqnarray}
D = P V \;,
\end{eqnarray}
with $P$ positive semidefinite. The unitary $V$ minimizes the norm distance from
$D$ yielding the inequality
\begin{eqnarray}
||D-V||_{2}  &=&  \sqrt{\sum_{k}\left[\sqrt{\lambda_{k}}-1\right]^{2}}
  \leqslant  \sqrt{\sum_{k}\left|\lambda_{k}-1\right|} \nonumber 
\\  \label{Q0} 
 & \leqslant & 
\sqrt{3}\; d_{C\bar{C}}\;(1-\eta_{0})^{1/4} \;, 
\end{eqnarray}
where we introduced  the eigenvalues $\lambda_{k}$
of  $D^{\dag}D$ and used Eq.~(\ref{IMPRT}) to bound them
according to the inequality $|\lambda_{k}-1|\leqslant3\; 
d_{C\bar{C}}\; \sqrt{1 -\eta_0}$
(in these expressions $d_{C\bar{C}} = d_{C}d_{\bar{C}}$ is the dimension
of the system  $C\bar{C}$ and $\| \Theta\|_2$ stands for  $\sqrt{ \sum_{kk^\prime} |\Theta_{kk^\prime}|^2}$ with $\Theta_{kk^\prime}$ being the matrix elements of the
operator $\Theta$).
The inequalities~(\ref{Q0}) and  (\ref{neweq})
show that for $L\rightarrow\infty$, $D$  can be
approximated arbitrarily well by the unitary $V$.
We can hence define $V^\dag$ as our decoding transformation which
inverts the mapping~(\ref{DEFD}) 
and transforms the ``bad'' representations $|\phi_k\rangle_M$ of the $|\psi_k\rangle_{C\bar{C}}$
into the ``good'' representations $|\psi_k\rangle_M$. It is worth stressing that,
by construction, $V$ does not depend upon the input state 
$|\psi\rangle_{C\bar{C}}$ of the system $C\bar{C}$. 

As mentioned in the introduction of this section, a similar procedure can
used to defined the encoding protocol of the uploading protocol. Without entering
into the details we simply notice that in this case $D$ and the vectors $|\phi_k\rangle_{M}$ will be defined by
replacing $W$ of Eq.~(\ref{eq:maink}) with the transformation $W^\prime$ of
Eq.~(\ref{wwwprime}). Taking the polar decomposition of such new $D$ it will 
yield the unitary $V^\prime$ which will be used as encoding for the uploading
scheme.

In the following section we will evaluate the transfer fidelities associated
with such a choice of decoding and encoding transformation, 
showing that they can arbitrarily increased by choosing
$L$ sufficiently high.

\subsection{Fidelity of the downloading protocol} \label{sec:fid}

Let $|\psi\rangle_{C\bar{C}} = \sum_k \alpha_k |\psi_k\rangle_{C\bar{C}}$ 
be a generic input state of $C\bar{C}$.
To evaluate the downloading fidelity $F_{\textrm{down}}$ associated with our decoding scheme
we need to compare the state of $M$ at the end of the protocol with
the state $|\psi\rangle_M = \sum_k \alpha_k |\psi_k\rangle_{M}$, i.e.
\begin{eqnarray}
F_{\textrm{down}}
(\psi)\;\equiv\;{}_{M}\langle\psi|V^{\dag}\; R_{M}\; V|\psi\rangle_{M}\;.\end{eqnarray}
Here $V^\dag$ is the decoding transformation defined in the previous section, and 
$R_{M}$ is the state of the memory after the application of the unitary $W$, 
i.e. \begin{eqnarray}
R_{M} & \equiv & \textrm{Tr}_{C\bar{C}}\left[W(|\psi\rangle_{C\bar{C}}\langle\psi|\otimes|0\rangle_{M}\langle0|)W^{\dag}\right]\nonumber \\
 & = & \eta\;|\phi\rangle_{M}\langle\phi|+(1-\eta)\;\sigma_{M}\;.\end{eqnarray}
 In the above expression we used Eqs.~(\ref{eq:main}) and~(\ref{impo})
and introduced the density matrix $\sigma_{M}\equiv
\textrm{Tr}_{\bar{C}}[|\Delta\rangle_{\bar{C}M}\langle\Delta|]$.
By linearity we get \begin{eqnarray}
F_{\textrm{down}}
(\psi) &=& \eta\;|{}_{M}\langle\phi|V|\psi\rangle_{M}|^{2}\nonumber \\
&&+ \; (1-\eta)\;{}_{M}\langle\psi|V^{\dag}\;\sigma_{M}\; V|\psi\rangle_{M}\nonumber \\
 & \geqslant & \eta\;|{}_{M}\langle\phi|V|\psi\rangle_{M}|^{2}\;.\label{fin1}\end{eqnarray}
We now bound the term on the right hand side as follows
 \begin{eqnarray}
|_{M}\langle\phi|V|\psi\rangle_{M}| & = & |_{M}\langle\phi|V-D+D|\psi\rangle_{M}|\label{fin2}\\
 & \geqslant & |_{M}\langle\phi|D|\psi\rangle_{M}|-|_{M}\langle\phi|D-V|\psi\rangle_{M}|\;,\nonumber \end{eqnarray}
and use the inequality~(\ref{Q0}) to write \begin{eqnarray*}
|_{M}\langle\phi|D-V|\psi\rangle_{M}|\leqslant||D-V||_{2}\leqslant\sqrt{3}\; d_{C\bar{C}}\;(1-\eta_{0})^{1/4}\;.\end{eqnarray*}
If $|\psi\rangle_{M}$ was a basis state $|\psi_{k}\rangle_{M},$
then $|_{M}\langle\phi|D|\psi\rangle_{M}|=1$ by the definition Eq.~(\ref{DEFD})
of $D$. For generic  $|\psi\rangle_{M}$ instead we can use the linearity
to find after some algebra that \begin{eqnarray}
\sqrt{\eta}\;|_{M}\langle\phi|D|\psi\rangle_{M}|\;\geqslant
\sqrt{\eta_{0}}\;-\;3\; d_{C\bar{C}}\;\sqrt{1-\eta_{0}}\;.\label{impo300}\end{eqnarray}
 Therefore Eq.~(\ref{fin2}) gives \begin{eqnarray}
\sqrt{\eta}\;|{}_{M}\langle\phi|V|\psi\rangle_{M}| & > & \sqrt{\eta_{0}}\;-5\; d_{C\bar{C}}\;(1-\eta_{0})^{1/4}\;,\label{fin4}\end{eqnarray}
which replaced in Eq.~(\ref{fin1}) yields
\begin{eqnarray}
F_{\textrm{down}}(\psi) & \geqslant & \eta_{0}\;-10\; d_{C\bar{C}}\;(1-\eta_{0})^{1/4}\;,
\label{fin1000}\end{eqnarray}
for all input states $|\psi\rangle_{C\bar{C}}$.
According to Eq.~(\ref{neweq}) it then follows that by choosing $L$ sufficiently
big our downloading protocol will yield transferring fidelities arbitrarily close to one.

\subsection{Fidelity of the uploading protocol}
Following the analysis of Sec.~(\ref{s:up}) 
the fidelity for uploading a state $|\psi\rangle_{M}$ into $\bar{C}C$
is given by \begin{eqnarray}
\lefteqn{F_{\text{up}}(\psi)\equiv}\\
 &  & _{C\bar{C}}\langle\psi|\textrm{Tr}_{M}\left[W^{\prime\dag} V^\prime
\left(|\psi\rangle_{M}\langle\psi|\otimes|0\rangle_{\bar{C}C}\langle0|\right)
{V}^{\prime\dag}W^\prime\right]|\psi\rangle_{C\bar{C}}.\nonumber \end{eqnarray}
 A lower bound for this quantity is obtained by replacing the trace
over $M$ with the expectation value on $|0\rangle_{M}$, i.e. \begin{eqnarray}
F_{\text{up}}(\psi) &\geqslant& 
{_{C\bar{C}}\langle}\psi|{}_{M}\langle0|W^{\prime\dag}V^\prime
\big(|\psi\rangle_{M}\langle\psi|\otimes|0\rangle_{\bar{C}C}\langle0|\big) \nonumber \\
&&\qquad \qquad \qquad\qquad \times\;
V^{\prime\dag}W^\prime|0\rangle_{M}|\psi\rangle_{C\bar{C}}\nonumber \\
 & = & \left|_{C\bar{C}}\langle0|{}_{M}\langle\psi|V^{\prime\dag}W^\prime
|0\rangle_{M}|\psi\rangle_{C\bar{C}}\right|^{2}\label{vvv} \\
 & = & \eta^\prime
\;\left|_{M}\langle\psi|V^{\prime\dag}|\phi\rangle_{M}\right|^{2}=\eta^\prime
\;\left|_{M}\langle\phi|V^\prime|\psi\rangle_{M}\right|^{2}. \nonumber 
\end{eqnarray}
In deriving this equation  we used Eq.~(\ref{tildewww}) and 
 a decomposition of the form
 of Eq.~(\ref{eq:main}) to simplify the vector $
W^\prime
|0\rangle_{M}|\psi\rangle_{C\bar{C}}$. In this case $\eta^\prime$
is defined as in Eq.~(\ref{eq:fid}) with $\tau$ being replaced by 
 $\tau^\prime$ of Eq.~(\ref{CPMAPprime}). Since we are assuming that
this CP map satisfies the condition~(\ref{condition20}) it follows that also
$\eta^\prime$ obeys an  inequality of the form (\ref{eq:fid11}) with $K$ and $\kappa$
replaced by new constants $K^{\prime}$ and $\kappa^{\prime}\in ]0,1[$.
We also notice that last term of Eq.~(\ref{vvv}) has the same form of 
the lower  bound~(\ref{fin1}) of the downloading fidelity. 
Therefore, by applying
the same derivation of the previous section we can write 
\begin{equation}
F_{\text{up}}(\psi) \geqslant\eta_{0}^\prime
\;-10\; d_{C\bar{C}}\;(1-\eta_{0}^\prime)^{1/4}\;,\label{fin3000}\end{equation}
with 
\begin{eqnarray}\label{neweqprime}
1 -\eta_0^\prime  \leqslant  K^\prime \;(L-1)^{d_{\bar{C}}}\;
(\kappa^\prime)^{L-1}.\end{eqnarray}
This shows that, as in the downloading case, also the uploading fidelity converges to unity in the limit of large~$L$.

\section{Efficiency of Cooling}\label{sec:cool}
In this section we 
provide some numerical estimation of the quantities $\eta$ of Eq.~(\ref{eq:fid})
which measure the probability of finding the state $\bar{C}$ in $|0\rangle_{\bar{C}}$.
As seen in the previous sections this is the fundamental parameter
to bound the fidelities of both the downloading and uploading protocol.
Moreover, given an initial state $|\psi\rangle_{C\bar{C}}$, $\eta$ measures
the success probability of ``cooling'' it down to the state $|0\rangle_{C\bar{C}}$
during the downloading process. 
According to Eq.~(\ref{eq:fid11}) the quantity  $\eta$ will \emph{asymptotically} converge
 exponentially fast to unity. However Eq.~(\ref{eq:fid11}) 
does not tell us from which point onwards the convergence is exponentially fast, 
so it would be nice to have alternative 
ways to estimate the convergence speed.

\begin{figure}[t]
\begin{centering}\includegraphics[width=1\columnwidth]{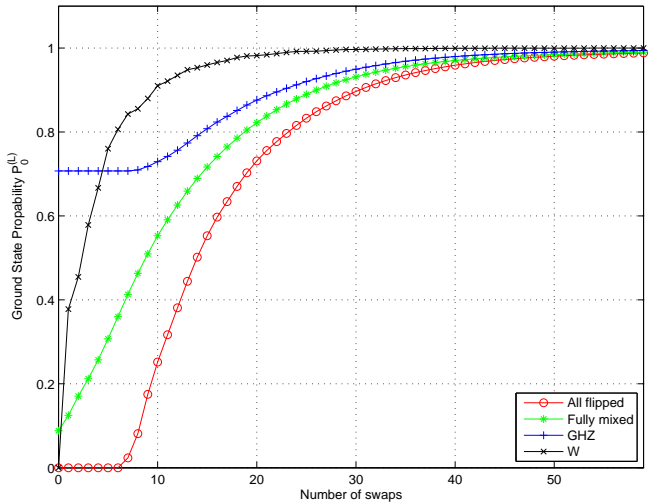}\par\end{centering}
\caption{\label{fig:swapping3}Convergence of the cooling protocol for the weighted graph 
of Fig. \ref{fig:memory} where the couplings among the spins 
is given by the Hamiltonian~(\ref{Heisenberg}) (the values of constants $d_{j,j^\prime}$ have been chosen
to be proportional to the length of the graph edge).
 Four different states $|\psi\rangle_{C\bar{C}}$
are considered: the fully flipped state $|1111111\rangle_{C\bar{C}},$
a GHZ state $(|0000000\rangle_{C\bar{C}}+|1111111\rangle_{C\bar{C}})/\sqrt{2},$
a fully mixed state $\rho_{C\bar{C}}=\openone_{C\bar{C}}/2^7$, and a W state
$\frac{1}{\sqrt{7}}(|1000000\rangle_{C\bar{C}}+|0100000\rangle_{C\bar{C}}+\cdots+|0000001\rangle_{C\bar{C}})$.}
\end{figure}

To simplify the analysis in the following, we will concentrate on 
the spin network model of Fig.~\ref{fig:memory} assuming a Heisenberg 
 Hamiltonian of the form
\begin{eqnarray}
H= \sum_{(j,j^\prime)\in G} d_{j,j^\prime}\big( X_j X_{j^\prime} + Y_j Y_{j^\prime} + Z_j Z_{j^\prime} \big) \;,
\label{Heisenberg}
\end{eqnarray}
which  conserves the total magnetization  
along the $z$ axis (here $X_j$, $Y_j$ and $Z_j$ are the Pauli operators of the $j$-th spin and
the summation is performed over all the edges of the weighted graph $G$ associated with the network). Moreover we will take  
the vector $|0\rangle_C$ to be the configuration where all the qubits of $C$ are in the spin-down state,
i.e. $|0\rangle_C \equiv |00\cdots 00\rangle_C$. For this choice
of the controller state our main assumption of ergodicity Eq.~(\ref{condition20})
is numerically found to be correct for the coupling graph depicted in Fig.~\ref{fig:memory}. The fixed point is given by $|0\rangle_{\bar{C}} \equiv |00\cdots 00\rangle_{\bar{C}}$ (more general conditions of ergodicity for Heisenberg models are given in~\cite{PAPER}). 
In this context 
$\eta$ coincides then with 
the probability $P_0^{(L)}$ of finding no excitations on the system
after $L$ steps of the protocol.  Some numerical examples showing the dependence of $\eta$ upon the initial state are presented in Fig.~\ref{fig:swapping3}. As expected, asymptotically  $P_0^{(L)}$ is seen to converge exponentially fast.

%%%%%%%%%%%%%%%%%%%%%%%%%%
%%%%%%%%%%%%%%%%%%%%%%%%%%%%%%%%%%%%%%
An approximate estimation of $P_{0}^{(L)}$ can be easily obtained 
by looking at the 
\emph{average} number of spin-up on $C\bar{C}$ after $L$ swaps, i.e. 
\begin{eqnarray}
\left\langle \hat{N}\right\rangle _{C\bar{C}}^{(L)} & \equiv & \textrm{Tr}_{C\bar{C}}
\left[ \hat{N}\; \rho^{(L)}_{C\bar{C}} \right]\;,
\end{eqnarray}
with $\rho_{C\bar{C}}^{(L)} \equiv \textrm{Tr}_M \big[ 
W \big(|\psi\rangle_{C\bar{C}}\langle 
\psi| \otimes |0\rangle_M\langle 0| \big) W^\dag\big]$ being 
the reduced density matrix of $C{\bar{C}}$ and 
with $\hat{N} \equiv  \sum_{k\in C,\bar{C}}\left(Z_{k} + 1\right)/2$
(here $Z_{k}$ is the $z$-Pauli matrix of the $k$-th spin).
These quantities are related by\begin{equation}
P_{0}^{(L)}\geqslant 1-\left\langle \hat{N}\right\rangle _{C\bar{C}}^{(L)}.\end{equation}
%%%%%%%%%%%%%%%%%%%%%%%%%%%%%%%%%%%%%%%%%%%%%%%%%%%
\begin{figure}[t]
\begin{centering}\includegraphics[width=1\columnwidth]{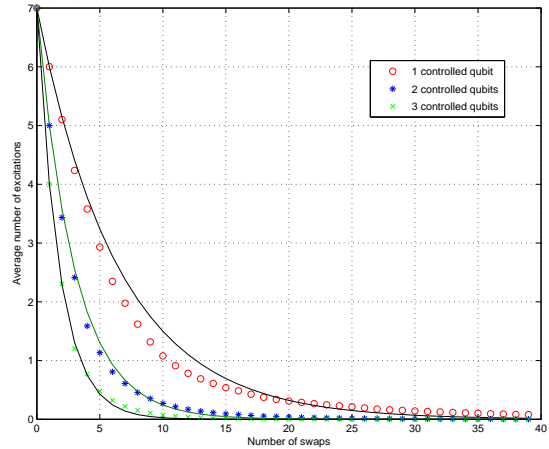}\par\end{centering}
\caption{\label{fig:swapping2}Comparison of the approximation Eq. (\ref{eq:appr})
with exact numerical results. Shown is the average number of excitations
$\left\langle \hat{N}\right\rangle _{C\bar{C}}$ on an open Heisenberg spin
chain with $7$ sites and equal couplings as a function of the number
of swaps to the memory. The initial state is taken to be $|1111111\rangle_{C\bar{C}}$,
i.e. with a maximal number of excitations. The three curves correspond
to different sizes of the region $|C|$ controlled by Alice, and the time interval $t$
has been chosen for each curve independently to fit the approximation given in Eq.~(\ref{eq:appr}).}
\end{figure}
%%%%%%%%%%%%%%%%%%%%%%%%%%%%%%%%%%%%%%%%%%%%%%%%%
%%%%%%%%%%%%%%%%%%%%%%%%%%%%%%%%%%%%%%%%%%%%%%%%%%%
\begin{figure}[b!]
\begin{centering}\includegraphics[width=1\columnwidth]{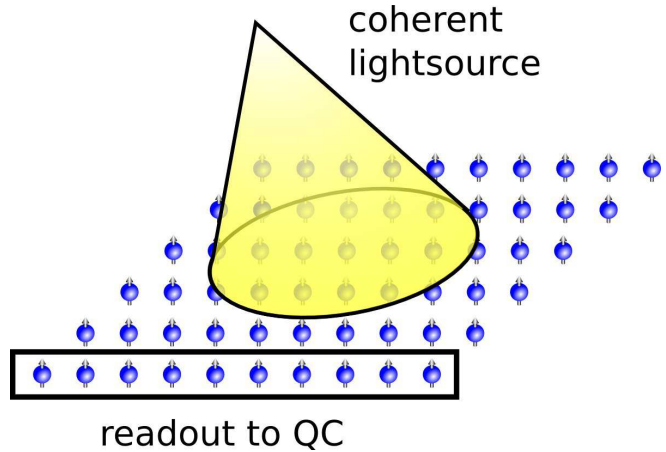}\par\end{centering}
\caption{\label{fig:ccd2}A CCD-like application of our protocol could allow a light sensitive array of qubits to be read out \emph{coherently} by a quantum computer without ``disturbing'' the qubits much.}
\end{figure}
%%%%%%%%%%%%%%%%%%%%%%%%%%%%%%%%%%%%%%%%%%%%%%%%%
To get an approximation for the average
number of excitations on the graph we assume now that the time interval
$t$ is chosen such that $U$ shuffles the excitations on the
graph in a fully random way. For specific systems and specific times intervals, 
this {}``classical'' behavior might not be true due to interferences,
but for general times it is a good 
approximation (see Fig.~\ref{fig:swapping2}). Let $|C|$ be the number of
edges on the graph controlled by Alice, and $|\bar{C}|$ the number
of uncontrolled edges. On average, each swap takes approximately a
ratio $|C|/(|C|+|\bar{C}|)$ of excitations from the graph to the
memory. We then get \begin{equation}
\left\langle \hat{N}\right\rangle _{C\bar{C}}^{(L)}\approx\left\langle \hat{N}\right\rangle _{C\bar{C}}^{(0)}\left(\frac{1}{1+\left|C\right|/\left|\bar{C}\right|}\right)^{L}\;.
\label{eq:appr}\end{equation}
This is a reasonable  result which shows that 
the fidelity depends on the initial
number of excitations and on the relative size of the controlled region
with respect to the uncontrolled region.

\section{Conclusion} \label{sec:con}

We have given an explicit protocol for controlling and cooling a large permanently
coupled system by accessing a small subsystem only. 
As we have shown, the applicability relies only on the invariant property~(\ref{condition20}) of a CPT map. Since we had to assume a large quantum memory in order to control the system, this protocol is not useful for replacing control in a homogeneous setup, but may well have applications in inhomogeneous scenarios (when control is harmful or expensive in some regions but easy in others). For example, we imagine a CCD-like application, in which a set of permanently coupled qubits is read out by a Quantum Computer in a coherent manner (see Fig.~\ref{fig:ccd2}).

\acknowledgments 
V.G. acknowledges the Quantum Information research program of
Centro di Ricerca Matematica Ennio De Giorgi of Scuola Normale
Superiore for financial support.

\appendix

\section{Evolution of $\bar{C}$}\label{appendix1}
Here we derive  the evolution~(\ref{NEWREF}) of the subset $\bar{C}$ in terms of the CP map~(\ref{CPMAP}).
First rewrite the reduced density matrix~(\ref{impo1}) 
as follows 
\begin{widetext}
\begin{eqnarray} \nonumber
\rho_{\bar{C}}^{(L)} &=&  \mbox{Tr}_{C M} 
\left[ W \left( |\Psi\rangle_{C\bar{C}}\langle \Psi| \otimes
 |0\rangle_M\langle 0|   \right) W^\dag \right]  \\
&=&   \mbox{Tr}_{C} \Big[   
\cdots \mbox{Tr}_{M_2}\Big[ S_2 U  \Big( \mbox{Tr}_{M_1} 
\Big[ S_1 U \Big( |\Psi\rangle_{C\bar{C}}\langle \Psi| \otimes
 |0\rangle_{M_1}\langle 0|   \Big) U^\dag S_1^\dag \Big] 
\otimes |0\rangle_{M_2}\langle 0| \Big)
U^\dag S_2^\dag \Big] \label{NEWEQ2}
\cdots  \Big]   \;.
\end{eqnarray}
For  the sake of clarity it is useful to explicitly denote the subsystems on which
the various operators are acting on (e.g. $\Theta_{AB}$ indicates that the
operator 
$\Theta$ acts non trivially only on the subsystems $A$ and $B$, while
it is the identity elsewhere). By doing so and by using the properties~\cite{NOTE2} of the swap 
it is easy to verify the following identities:
\begin{eqnarray}
&\mbox{Tr}_{M_\ell}\Big[ S_{\ell} U  \Big( 
\rho_{\bar{C}} \otimes |0\rangle_{C}\langle 0|
\otimes |0\rangle_{M_\ell}\langle 0| \Big)
U^\dag S_{\ell}^\dag \Big] 
= \mbox{Tr}_{M_\ell}\Big[ S_{C M_\ell} U_{C\bar{C}}  \Big( 
\rho_{\bar{C}}\otimes |0\rangle_{C}\langle 0|
\otimes |0\rangle_{M_\ell}\langle 0| \Big)
U_{C\bar{C}}^\dag S_{C M_\ell}^\dag \Big]
& \nonumber  \\
&= \mbox{Tr}_{M_\ell}\Big[ S_{C M_\ell} U_{C\bar{C}}  
\Big( S_{C M_\ell}^\dag S_{C M_\ell}\Big) \Big( 
\rho_{\bar{C}}\otimes |0\rangle_{C}\langle 0|
\otimes |0\rangle_{M_\ell}\langle 0| \Big) \Big(S_{C M_\ell}^\dag S_{C M_\ell} \Big)
U_{C\bar{C}}^\dag S_{C M_\ell}^\dag \Big]& \nonumber \\
&= \mbox{Tr}_{M_\ell}\Big[  U_{M_\ell\bar{C}}    \Big( 
\rho_{\bar{C}} \otimes |0\rangle_{M_\ell}\langle 0|
\otimes |0\rangle_{C}\langle 0| \Big)
U_{M_\ell\bar{C}}^\dag  \Big] =  \tau( 
\rho_{\bar{C}} )\otimes |0\rangle_{C}\langle 0| \;, &
\end{eqnarray}
which holds for all $\rho_{\bar{C}}$ and $\ell $. Equation~(\ref{NEWREF}) then follows by  replacing this into Eq.~(\ref{NEWEQ2}) for all $\ell >2$ and 
by using the identity
\begin{eqnarray}
&&\mbox{Tr}_{M_1} \Big[ S_1 U\Big( |\Psi\rangle_{C\bar{C}}\langle \Psi| \otimes
 |0\rangle_{M_1}\langle 0|   \Big) U^\dag  S_1^\dag \Big]  =  
 \mbox{Tr}_{M_1} \Big[ S_{CM_1} U_{C\bar{C}} \Big( |\Psi\rangle_{C\bar{C}}\langle \Psi| \otimes
 |0\rangle_{M_1}\langle 0|   \Big) U^\dag_{C\bar{C}}  S_{CM_1}^\dag \Big]  \\
&& =  \mbox{Tr}_{M_1} \Big[  U_{M_1\bar{C}}  \Big( |\Psi\rangle_{M_1\bar{C}}\langle \Psi| \otimes
 |0\rangle_{C}\langle 0|   \Big) U^\dag_{M_1\bar{C}}   \Big] =   \mbox{Tr}_{M_1} \Big[  U_{M_1\bar{C}}  \Big( |\Psi\rangle_{M_1\bar{C}}\langle \Psi|  \Big) U^\dag_{M_1\bar{C}}   \Big]  \otimes
 |0\rangle_{C}\langle 0|  \equiv \rho_{\bar{C}}^\prime \otimes
 |0\rangle_{C}\langle 0|  \;, \nonumber 
\end{eqnarray}
\end{widetext}
with $\rho_{\bar{C}}^\prime$ as in Eq.~(\ref{newref1}).

\section{Decomposition equations} \label{appendix2}

Here we give a decomposition of the state after applying
the $W$ operator of Eq.~(\ref{www}). This
will allow us to estimate the fidelities for state
transfer in terms of the relaxing properties of the map $\tau.$ 

Let
$|\psi\rangle_{C\bar{C}}\in\mathcal{H}_{C\bar{C}}$ be an arbitrary
state. We notice that the $C$ component of $W|\psi\rangle_{C\bar{C}}|0\rangle_{M}$
is always $|0\rangle_{C}$. Therefore we can decompose it as follows
\begin{equation}
W|\psi\rangle_{C\bar{C}}|0\rangle_{M}=|0\rangle_{C}\otimes\left[\sqrt{\eta}|0\rangle_{\bar{C}}|\phi\rangle_{M}+\sqrt{1-\eta}|\Delta\rangle_{\bar{C}M}\right]\label{eq:main}\end{equation}
with $\eta\in [0,1]$ and
 with $|\Delta\rangle_{\bar{C}M}$ being a normalised vector of $\bar{C}$
and $M$ which satisfies the identity \begin{eqnarray}
_{\bar{C}}\langle0|\Delta\rangle_{\bar{C}M}=0\;.\label{impo}\end{eqnarray}
 It is worth stressing that in the above expression $\eta$, $|\phi\rangle_{M}$
and $|\Delta\rangle_{\bar{C}M}$ are depending on $|\psi\rangle_{C\bar{C}}$.
In a similar way we can decompose the vector obtained by acting with $W^{\dag}$ 
on the first term of Eq.~(\ref{eq:main}), i.e.
\begin{equation}
W^{\dag}|0\rangle_{C\bar{C}}|\phi\rangle_{M}=\sqrt{\tilde{\eta}}\;|\psi
\rangle_{C\bar{C}}|0\rangle_{M}+\sqrt{1-\tilde{\eta}}\;|
\tilde{\Delta}\rangle_{C\bar{C}M},\label{eq:main2}
\end{equation}
where $|\tilde{\Delta}\rangle_{C\bar{C}M}$ is the orthogonal complement
of $|\psi\rangle_{C\bar{C}}|0\rangle_{M},$ i.e. \begin{equation}
_{\bar{C}C}\langle\psi|{}_{M}\langle0|\tilde{\Delta}\rangle_{C\bar{C}M}=0\;.\label{eq:o2}\end{equation}
Multiplying Eq. (\ref{eq:main2}) from the left with $_{C\bar{C}}\langle\psi|_{M}\langle0|$
and using the conjugate of Eq. (\ref{eq:main}) we find that $\eta=\tilde{\eta}.$
An expression of $\eta$ in terms of $\tau$ can be obtained by using Eq.~(\ref{impo1}).
Therefore from Eq.~(\ref{eq:main}) and the orthogonality 
relation~(\ref{impo}) it follows that 
\begin{equation}
\eta={}_{\bar{C}}\langle0|\tau^{L-1}\left(\rho_{\bar{C}}^{\prime}\right)
|0\rangle_{\bar{C}},\label{eq:fid}\end{equation}
 which, since $\tau$ is relaxing, shows that $\eta\rightarrow1$ for
$L\rightarrow\infty$. Moreover we can use \cite{TERHAL} to claim
that \begin{eqnarray}
1-\eta & = & |{}_{\bar{C}}\langle0|\tau^{L-1}\left(\rho_{\bar{C}}^{\prime}\right)|0\rangle_{\bar{C}}-1|\nonumber \\
 & \leqslant & \|\tau^{L-1}\left(\rho_{\bar{C}}^{\prime}\right)-|0\rangle_{\bar{C}}\langle0|\|_{1}\nonumber \\ \label{eq:fid11}
 & \leqslant & K \;(L-1)^{d_{\bar{C}}}\;\kappa^{L-1},\end{eqnarray}
 where  $\| \Theta \|_1 = \sqrt{\textrm{Tr}[ \Theta^\dag\Theta]}$ is the trace norm 
of the operator $\Theta$,
$K$ is a constant which depends upon $d_{\bar{C}}\equiv\mbox{dim}\mathcal{H}_{\bar{C}}$,
and where $\kappa\in]0,1[$ is the second largest of the moduli of
eigenvalues of $\tau.$


\begin{thebibliography}{10}
\bibitem{NOTE2}
Given two systems  $A$ and $B$ of dimenstion $d$, characterized by
canonical basis $\{ |1\rangle_{A}, \cdots, |d\rangle_A \}$ and  
$\{ |1\rangle_{B}, \cdots, |d\rangle_B \}$, respectively,  
the swap operators introduced in Sec.~\ref{s:down} 
is the unitary transformation defined as
$S_{AB} = S_{AB}^\dag \equiv \sum_{j=1}^d \sum_{k=1}^d |j\rangle_A \langle k| 
\otimes | k\rangle_B \langle j|$.
When applied to a joint state $|\Psi\rangle_{AB}$ 
it exchanges the degree of freedom of $A$ with the degree 
of freedom of $B$. More generally, for any  $\Theta_A$ and $\Omega_A$
 operators of $A$, and $\Theta_B$ and $\Omega_B$ 
their corresponding conterparts on $B$, we have
$S_{AB} \;  \left( \Theta_A \otimes 
\Omega_B \right) \; S_{AB} = \Omega_A\otimes \Theta_B$.
\bibitem{NOTE3}
The condition of Eq.~(\ref{condition20}) is compatible with  $|0\rangle_{C\bar{C}}$ 
being an eigenvector of $H$. In particular we can assume it to be the
ground state of the system.
\bibitem{notezero}
We use the adjoint $V^\dag$  to indicate the decoding operator instead of $V$,
to maintain the same notation of Ref.~\cite{PAPER}.
\end{thebibliography}
\end{document}